\documentclass[twocolumn,showpacs,prc,floatfix]{revtex4}
\usepackage{graphicx}
\usepackage{amsmath}
\newcommand{\bs}{\begin{sloppypar}} \newcommand{\es}{\end{sloppypar}}
\newcommand{\p}[1]{(\ref{#1})}
\newcommand\beq{\begin{eqnarray}} \newcommand\eeq{\end{eqnarray}}
\newcommand\beqstar{\begin{eqnarray*}} \newcommand\eeqstar{\end{eqnarray*}}
\newcommand{\beqe}{\begin{equation}}
\newcommand{\beqestar}{\begin{equation*}}
\newcommand{\bal}{\begin{align}}

\begin {document}
\title{Antiferromagnetic spin phase
transition \protect\\  in nuclear matter with effective Gogny
interaction }
\author{ A. A. Isayev}
 \affiliation{Kharkov Institute of
Physics and Technology, Academicheskaya Street 1,
 Kharkov, 61108, Ukraine}
 \author{J. Yang}
 \affiliation{Dept. of Physics and Center for Space Science and Technology,\\
Ewha Womans University, Seoul 120-750, Korea\\ and  Center for
High Energy Physics, Kyungbook National University, Daegu 702-701,
Korea}
\begin{abstract}   The possibility of ferromagnetic and antiferromagnetic
phase transitions in symmetric nuclear matter is analyzed within
the framework of a Fermi liquid theory with  the effective Gogny
interaction. It is shown
 that at some critical density nuclear
matter with  the D1S effective force undergoes a phase transition
to the antiferromagnetic spin state (opposite directions of
neutron and proton spins). The self--consistent equations of spin
polarized nuclear matter with the D1S force have no solutions
corresponding to  ferromagnetic spin ordering (the same direction
of neutron and proton spins) and, hence, the ferromagnetic
transition does not appear. The dependence of the
antiferromagnetic spin polarization parameter as a function of
density  is found at zero temperature.
\end{abstract}
\pacs{21.65.+f; 75.25.+z; 71.10.Ay} \maketitle

\section{Introduction}
 The spontaneous appearance of  spin polarized states in nuclear
matter is a topic of a great current interest due to its relevance
in astrophysics. In particular, the effects of spin correlations
in the medium strongly influence the neutrino cross section and,
hence,  neutrino mean free path. Therefore, depending on whether
nuclear matter is spin polarized or not, drastically different
scenarios of supernova explosion and cooling of neutron stars can
be realized. Another aspect relates to pulsars, which are
considered to be rapidly rotating neutron stars, surrounded by
strong magnetic field. There is still no general consensus
regarding the mechanism to generate such a strong magnetic field
of a neutron star. One of the hypotheses is that a magnetic field
can be produced by a spontaneous ordering of spins in the dense
stellar core.

 The
possibility of a phase transition of normal neutron and nuclear
matter to the ferromagnetic spin state was studied by many
authors~\cite{R}--\cite{TT}, predicting the ferromagnetic
transition at $\varrho\approx(2$--$4)\varrho_0$ for different
parametrizations of Skyrme forces
($\varrho_0=0.16\,\mbox{fm}^{-3}$ is the nuclear matter saturation
density).  In particular, the stability of strongly asymmetric
nuclear matter with respect to spin fluctuations was investigated
in Ref.~\cite{KW}, where it was shown that  the system with
localized protons can develop a spontaneous polarization, if  the
neutron--proton spin interaction exceeds some threshold value.
This conclusion was confirmed also by calculations within the
relativistic Dirac--Hartree--Fock approach to strongly asymmetric
nuclear matter~\cite{BMNQ}. Competition between ferromagnetic (FM)
and antiferromagnetic (AFM) spin ordering in symmetric nuclear
matter with the Skyrme effective interaction was studied in
Ref.~\cite{I}, where it was clarified that the FM spin state is
thermodynamically preferable to the AFM one for all relevant
densities. However,  strongly asymmetric nuclear matter with
Skyrme forces undergoes a phase transition to a state with
oppositely directed spins of neutrons and protons~\cite{IY}.

For the models with realistic nucleon--nucleon (NN) interaction,
the ferromagnetic phase transition seems to be suppressed up to
densities well above $\varrho_0$~\cite{PGS}--\cite{H}. In
particular, no evidence of ferromagnetic instability has been
found in recent studies of neutron matter~\cite{VPR} and
asymmetric nuclear matter~\cite{VB} within the
Brueckner--Hartree--Fock approximation with realistic Nijmegen II,
Reid93, and Nijmegen NSC97e NN interactions. The same conclusion
was obtained in Ref.~\cite{FSS}, where the magnetic susceptibility
of neutron matter was calculated with the use of the Argonne
$v_{18}$ two--body potential and Urbana IX three--body potential.

Thus, the issue of appearance of spin polarized states in nuclear
matter is a controversial one and models with  effective Skyrme
and realistic NN potentials predict different results. From this
point of view, it is interesting to attract another type of NN
interaction and to compare the results for this   NN potential
with the previous results. Here we continue the study of spin
polarizability of nuclear matter with the use of an effective NN
interaction, namely, we utilize the  effective Gogny
force~\cite{DG,BGG}. In addition, the reason to choose the Gogny
interaction is as follows. It is known that the Skyrme interaction
is a density dependent zero range NN potential. Its attractive
advantage is relative simplicity and successfulness in describing
nuclei and their excited states. However, in many cases the finite
range part of the nuclear interaction has the same importance as
its density dependent zero range part~\cite{ZSSL,BL}. This
disadvantage of the Skyrme interaction is overcome by the Gogny
interaction due to its finite range character. We will  find the
phase diagram of spin polarized nuclear matter with the Gogny
interaction and will compare it with the results for the Skyrme
and realistic NN potentials.

 As a framework of
consideration, we choose a Fermi liquid (FL) description of
nuclear matter~\cite{AKP}--\cite{AIPY}.  We explore the
possibility of FM and
 AFM phase transitions in nuclear matter, when
the spins of protons and neutrons are aligned in the same
direction or in the opposite direction, respectively. In contrast
to the approach, based on the calculation of magnetic
susceptibility, we obtain the self--consistent equations for the
FM and  AFM spin order parameters and solve them at zero
temperature. This allows us not only to determine  the critical
density of instability with respect to spin fluctuations, but also
to establish the density dependence of the order parameters and to
clarify the question of thermodynamic stability of various phases.

Note that we consider  the thermodynamic properties of spin
polarized states in nuclear matter up to the high density region
relevant for astrophysics. Nevertheless, we take into account the
nucleon degrees of freedom only,  although other degrees of
freedom, such as pions, hyperons, kaons,  or quarks could be
important at such high densities.
\section{Basic Equations}

 The normal states of nuclear matter are described
  by the normal distribution function of nucleons $f_{\kappa_1\kappa_2}=\mbox{Tr}\,\varrho
  a^+_{\kappa_2}a_{\kappa_1}$, where
$\kappa\equiv({\bf{p}},\sigma,\tau)$, ${\bf p}$ is the momentum,
$\sigma(\tau)$ is the projection of spin (isospin) on the third
axis, and $\varrho$ is the density matrix of the system. The
self-consistent matrix equation for determining the distribution
function $f$ follows from the minimum condition of the
thermodynamic potential \cite{AKP} and is
  \begin{eqnarray}
 f=\left\{\mbox{exp}(Y_0\varepsilon+
Y_4)+1\right\}^{-1}\equiv
\left\{\mbox{exp}(Y_0\xi)+1\right\}^{-1}.\label{2}\end{eqnarray}
Here the nucleon single particle energy $\varepsilon$ is defined
through the energy functional of the system $E(f)$ as
 \begin{eqnarray}
\varepsilon_{\kappa_1\kappa_2}(f)=\frac{\partial E(f)}{\partial
f_{\kappa_2\kappa_1}}. \label{1} \end{eqnarray} In Eq.~\p{2}, the
quantities $\varepsilon$ and $Y_4$ are matrices in the space of
$\kappa$ variables, with
$Y_{4\kappa_1\kappa_2}=Y_{4\tau_1}\delta_{\kappa_1\kappa_2}$
$(\tau_1=n,p)$, $Y_0=1/T,\ Y_{4n}=-\mu_n/T$ and $Y_{4p}=-\mu_p/T$
being  the Lagrange multipliers, $\mu_n$ and $\mu_p$  being the
chemical potentials of  neutrons and protons, and $T$ being   the
temperature.


Taking into account the possibility of FM and AFM phase
transitions in nuclear matter, the normal distribution function
$f$ of nucleons and  single particle energy $\varepsilon$ can be
expanded in the Pauli matrices $\sigma_i$ and $\tau_k$ in spin and
isospin
spaces
\begin{align} f({\bf p})&= f_{00}({\bf
p})\sigma_0\tau_0+f_{30}({\bf p})\sigma_3\tau_0\label{7.2}\\
&\quad + f_{03}({\bf p})\sigma_0\tau_3+f_{33}({\bf
p})\sigma_3\tau_3. \nonumber\\
 \varepsilon({\bf p})&=
\varepsilon_{00}({\bf
p})\sigma_0\tau_0+\varepsilon_{30}({\bf p})\sigma_3\tau_0\label{7.3}\\
&\quad + \varepsilon_{03}({\bf
p})\sigma_0\tau_3+\varepsilon_{33}({\bf p})\sigma_3\tau_3.
\nonumber
\end{align}
Expressions for  the distribution functions
$f_{00},f_{30},f_{03},f_{33}$
 in
terms of the quantities $\varepsilon$ read \begin{align}
f_{00}&=\frac{1}{4}\{n(\omega_{+,+})+n(\omega_{+,-})+n(\omega_{-,+})+n(\omega_{-,-})
\},\nonumber
 \\
f_{30}&=\frac{1}{4}\{n(\omega_{+,+})+n(\omega_{+,-})-n(\omega_{-,+})-n(\omega_{-,-})
\},\label{2.4}\\
f_{03}&=\frac{1}{4}\{n(\omega_{+,+})-n(\omega_{+,-})+n(\omega_{-,+})-n(\omega_{-,-})
\},\nonumber\\
f_{33}&=\frac{1}{4}\{n(\omega_{+,+})-n(\omega_{+,-})-n(\omega_{-,+})+n(\omega_{-,-})
\}.\nonumber
 \end{align} Here $n(\omega)=\{\exp(\omega/T)+1\}^{-1}$ and
\begin{gather}
\omega_{+,+}=\xi_{00}+\xi_{30}+\xi_{03}+\xi_{33},\;\nonumber\\
\omega_{+,-}=\xi_{00}+\xi_{30}-\xi_{03}-\xi_{33},\;\label{2.5}\\
\omega_{-,+}=\xi_{00}-\xi_{30}+\xi_{03}-\xi_{33},\;\nonumber\\
\omega_{-,-}=\xi_{00}-\xi_{30}-\xi_{03}+\xi_{33},\;\nonumber\end{gather}
where \begin{align*}\xi_{00}&=\varepsilon_{00}-\mu_{00},\;
\xi_{30}=\varepsilon_{30},\;
\\
\xi_{03}&=\varepsilon_{03}-\mu_{03},\;\xi_{33}=\varepsilon_{33},\\
\mu_{00}&={\frac{\mu_n+\mu_p}{2}},\quad
\mu_{03}={\frac{\mu_n-\mu_p}{2}}.\end{align*}
The quantity $\omega_{\pm,\pm}$, being the exponent in the Fermi
distribution function $n$, plays the role of the quasiparticle
spectrum. In the general  case,  the spectrum is fourfold split
due to the spin and isospin dependence of the single particle
energy $\varepsilon({\bf p})$ in Eq.~\p{7.3}. The branches
$\omega_{\pm,+}$ correspond to neutrons with spin up and spin
down, and  the branches $\omega_{\pm,-}$ correspond to protons
with spin up and spin down.

The distribution functions $f$ should satisfy the norma\-lization
conditions
\begin{align} \frac{4}{\cal
V}\sum_{\bf p}f_{00}({\bf p})&=\varrho,\label{3.1}\\
\frac{4}{\cal V}\sum_{\bf p}f_{03}({\bf
p})&=\varrho_n-\varrho_p\equiv\alpha\varrho,\label{3.3}\\
\frac{4}{\cal V}\sum_{\bf p}f_{30}({\bf
p})&=\varrho_\uparrow-\varrho_\downarrow\equiv\Delta\varrho_{\uparrow\uparrow},\label{3.2}\\
\frac{4}{\cal V}\sum_{\bf p}f_{33}({\bf
p})&=(\varrho_{n\uparrow}+\varrho_{p\downarrow})-
(\varrho_{n\downarrow}+\varrho_{p\uparrow})\equiv\Delta\varrho_{\uparrow\downarrow}.\label{3.4}
 \end{align}
 Here $\alpha$ is the isospin asymmetry parameter, $\varrho_{n\uparrow},\varrho_{n\downarrow}$ and
 $\varrho_{p\uparrow},\varrho_{p\downarrow}$ are the neutron and
 proton number densities with spin up and spin down,
 respectively;
 $\varrho_\uparrow=\varrho_{n\uparrow}+\varrho_{p\uparrow}$ and
$\varrho_\downarrow=\varrho_{n\downarrow}+\varrho_{p\downarrow}$
are the nucleon densities with spin up and spin down. The
quantities $\Delta\varrho_{\uparrow\uparrow}$ and
$\Delta\varrho_{\uparrow\downarrow}$ play the roles of  FM and AFM
spin order parameters~\cite{IY}.

In order to characterize spin ordering in the neutron and  proton
subsystems, it is convenient to introduce   neutron and proton
spin polarization parameters \beqe
\Pi_n=\frac{\varrho_{n\uparrow}-\varrho_{n\downarrow}}{\varrho_n},\quad
\Pi_p=\frac{\varrho_{p\uparrow}-\varrho_{p\downarrow}}{\varrho_p}.
\end{equation}

The self--consistent equations for the components of the single
particle energy have the form~\cite{I,IY} \bal\xi_{00}({\bf
p})&=\varepsilon_{0}({\bf p})+\tilde\varepsilon_{00}({\bf
p})-\mu_{00},\;
\xi_{30}({\bf p})=\tilde\varepsilon_{30}({\bf p}),\label{14.2} \\
\xi_{03}({\bf p})&=\tilde\varepsilon_{03}({\bf p})-\mu_{03}, \;
\xi_{33}({\bf p})=\tilde\varepsilon_{33}({\bf
p}).\nonumber\end{align} Here $\varepsilon_0({\bf
p})=\frac{\hbar^2{\bf p}^2}{2m_{0}}$ is the free single particle
spectrum, $m_0$ is the bare mass of a nucleon, and
$\tilde\varepsilon_{00},\tilde\varepsilon_{30},\tilde\varepsilon_{03},\tilde\varepsilon_{33}$
are the FL corrections to the free single particle spectrum,
related to the normal FL amplitudes $U_0({\bf k}),...,U_3({\bf k})
$ by formulas
\begin{align}\tilde\varepsilon_{00}({\bf
p})&=\frac{1}{2\cal V}\sum_{\bf q}U_0({\bf k})f_{00}({\bf
q}),\;{\bf k}=\frac{{\bf p}-{\bf q}}{2}, \label{14.1}\\
\tilde\varepsilon_{30}({\bf p})&=\frac{1}{2\cal V}\sum_{\bf
q}U_1({\bf k})f_{30}({\bf q}),\nonumber\\ 
\tilde\varepsilon_{03}({\bf p})&=\frac{1}{2\cal V}\sum_{\bf
q}U_2({\bf k})f_{03}({\bf q}), \nonumber\\
\tilde\varepsilon_{33}({\bf p})&=\frac{1}{2\cal V}\sum_{\bf
q}U_3({\bf k})f_{33}({\bf q}). \nonumber
\end{align}

Further we do not take into account the effective tensor forces,
which lead to coupling of the momentum and spin degrees of freedom
\cite{HJ,D,FMS}, and, correspondingly, to anisotropy in the
momentum dependence of FL amplitudes with respect to the spin
polarization axis.

   To obtain
 numerical results, we  use the  effective Gogny interaction. The
 amplitude of the NN interaction in this case reads
\bal \hat v({\bf p},{\bf q})&=t_0(1+x_0\hat
P_\sigma)\varrho^\gamma+\pi^{3/2}\sum_{i=1}^{2}\mu_i^3(W_i+B_i\hat
P_\sigma\label{10.2}\\
&\quad-H_i \hat P_\tau-M_i\hat P_\sigma \hat P_\tau)e^{-({\bf
p}-{\bf q})^2\mu_i^2/4}, \nonumber\end{align} where $\hat
P_\sigma$ and $\hat P_\tau$ are the spin and isospin exchange
operators, and $t_0,x_0,\mu_i,W_i,B_i,H_i$ and $M_i$ are some
phenomenological constants, characterizing a given parametrization
of the Gogny forces. In numerical calculations  we
  shall utilize the    D1S  potential~\cite{BGG}.
  Using the same procedure as in Ref.~\cite{AIP}, it is possible to
  find expressions for the normal FL amplitudes in terms of Gogny
  force parameters
   \begin{align}
U_0({\bf k})&=6t_0\varrho^\gamma+
2\pi^{3/2}\sum_{i=1}^{2}\mu_i^3(2B_i-2H_i-M_i+4W_i)\label{10.3}\\
&\quad -2\pi^{3/2}\sum_{i=1}^{2}e^{-{\bf
k}^2\mu_i^2}\mu_i^3(2B_i-2H_i-4M_i+W_i),\nonumber\\
U_1({\bf k})&=-2t_0\varrho^\gamma(1-2x_0)+
2\pi^{3/2}\sum_{i=1}^{2}\mu_i^3(2B_i-M_i)\nonumber\\
&\quad +2\pi^{3/2}\sum_{i=1}^{2}e^{-{\bf
k}^2\mu_i^2}\mu_i^3(2H_i-W_i),\nonumber\\
U_2({\bf k})&=-2t_0\varrho^\gamma(1+2x_0)-
2\pi^{3/2}\sum_{i=1}^{2}\mu_i^3(2H_i+M_i)\nonumber\\
&\quad -2\pi^{3/2}\sum_{i=1}^{2}e^{-{\bf
k}^2\mu_i^2}\mu_i^3(2B_i+W_i),
\nonumber\\
U_3({\bf k})&=-2t_0\varrho^\gamma \nonumber\\
&\quad- 2\pi^{3/2}\sum_{i=1}^{2}\mu_i^3M_i
-2\pi^{3/2}\sum_{i=1}^{2}e^{-{\bf k}^2\mu_i^2}\mu_i^3W_i.
\nonumber
\end{align}
 Thus,
with account of  expressions \p{2.4} for the distribution
functions $f$, we obtain the self--consistent equations \p{14.2},
\p{14.1}
 for the components of the single particle energy $\xi_{00}({\bf
p}),\xi_{30}({\bf p}),\xi_{03}({\bf p}),\xi_{33}({\bf p})$, which
should be solved jointly with the normalization conditions
\p{3.1}--\p{3.4}, determining the chemical potentials
$\mu_{00},\mu_{03}$,
 FM  and AFM spin
 order parameters
$\Delta\varrho_{\uparrow\uparrow}$,
$\Delta\varrho_{\uparrow\downarrow}$. Since the FL amplitudes in
Eqs.~\p{10.3} contain two Gaussian terms, the self--consistent
equations represent, in fact, the set of coupled integral
equations, which can be solved iteratively using the Gaussian mesh
points in the momentum space.

\section{Phase transitions in symmetric nuclear matter}
Early research on spin polarizability  of nuclear matter  was
based on the calculation of magnetic susceptibility and finding
its pole structure~\cite{VNB,RPLP}, determining the onset of
instability with respect to spin fluctuations. Here we shall solve
directly the self--consistent equations for the FM and AFM spin
order parameters  at zero temperature. In this study  we consider
the case of   symmetric nuclear matter ($\varrho_n=\varrho_p$).

The FM spin ordering corresponds to the case
$\Delta\varrho_{\uparrow\uparrow}\not=0,\xi_{30} ({\bf
p})\not=0,\Delta\varrho_{\uparrow\downarrow}=0,\xi_{33} ({\bf
p})=0$, and there are two unknown parameters
$\mu_{00},\Delta\varrho_{\uparrow\uparrow}$ and two unknown
functions $\xi_{00} ({\bf p}),\xi_{30} ({\bf p})$ ($\mu_{03}=0,
\varepsilon_{03} ({\bf p})=0$ as a consequence of isospin
symmetry). The AFM spin ordering corresponds to the case
$\Delta\varrho_{\uparrow\downarrow}\not=0,\xi_{33} ({\bf
p})\not=0,\Delta\varrho_{\uparrow\uparrow}=0,\xi_{30} ({\bf p})=0$
and we should find two  unknown parameters
$\mu_{00},\Delta\varrho_{\uparrow\downarrow}$ and two unknown
functions $\xi_{00} ({\bf p}),\xi_{33} ({\bf p})$.

In the FM spin state of symmetric nuclear matter we have
$\varrho_{n\uparrow}=\varrho_{p\uparrow},
\varrho_{n\downarrow}=\varrho_{p\downarrow}$ and nucleons with
spin up fill  the Fermi surface of radius $k_2$ and nucleons with
spin down  occupy the Fermi surface of radius $k_1$, which satisfy
the relationships \beqe
\frac{1}{3\pi^2}(k_2^3-k_1^3)=\Delta\varrho_{\uparrow\uparrow},\quad
\frac{1}{3\pi^2}(k_1^3+k_2^3)=\varrho.\end{equation}

Since at zero temperature there are no spin up nucleons with
$k>k_2$ and there are no spin down  nucleons with $k>k_1$, then,
as follows from Eq.~\p{2.5},
$\omega_{+,+}(k_2)=\omega_{+,-}(k_2)=0$,
$\omega_{-,+}(k_1)=\omega_{-,-}(k_1)=0$.

 In
the AFM spin state $\varrho_{n\uparrow}=\varrho_{p\downarrow},
\varrho_{n\downarrow}=\varrho_{p\uparrow}$ and neutrons with spin
up and protons with spin down   fill the Fermi surface of radius
$k_2$ and neutrons with spin down and protons with spin up occupy
the Fermi surface of radius $k_1$, satisfying the equations \beqe
\frac{1}{3\pi^2}(k_2^3-k_1^3)=\Delta\varrho_{\uparrow\downarrow},\quad
\frac{1}{3\pi^2}(k_1^3+k_2^3)=\varrho.\end{equation}

At zero temperature there are no spin up neutrons and spin down
protons with $k>k_2$ and there are no spin down  neutrons and spin
up protons with $k>k_1$. Hence, as follows from Eq.~\p{2.5},
$\omega_{+,+}(k_2)=\omega_{-,-}(k_2)=0$,
$\omega_{+,-}(k_1)=\omega_{-,+}(k_1)=0$.

In the totally FM polarized state we have $
\Delta\varrho_{\uparrow\uparrow}=\varrho,\;
k_2=k_F\equiv(3\pi^2\varrho)^{1/3}$ and the degrees of freedom,
corresponding  to nucleons with spin down are frozen ($k_1=0$).
For totally AFM polarized nuclear matter we have $
\Delta\varrho_{\uparrow\downarrow}=\varrho$, $k_2=k_F$, where
$k_2$ is given by the same expression  as in the FM case, since
now the degrees of freedom, corresponding to  neutrons with spin
down and protons with spin up, are frozen ($k_1=0$).

Now we present the results of the numerical  solution of the
self--consistent equations with the  D1S  Gogny effective force.
The main qualitative feature is that for the  D1S force there are
solutions corresponding to  AFM spin ordering and  there are no
solutions corresponding to  FM spin ordering. The reason is that
the sign of the multiplier $t_0(2x_0-1)$ in the density dependent
term of the FL amplitude $U_1$, determining spin--spin
correlations, is positive, and, hence, the corresponding term
increases with  increase of nuclear matter density, preventing
instability with respect to spin fluctuations. Contrarily, the
density dependent term $-2t_0\varrho^\gamma$ in the FL amplitude
$U_3$, describing spin--isospin correlations, is negative, leading
to spin instability with  oppositely directed spins of neutrons
and protons at higher densities. Here the situation is similar to
that with the Skyrme effective forces SLy4 and SLy5 in strongly
asymmetric nuclear matter~\cite{IY}, where analogous behavior of
the FL amplitudes $U_1$ and $U_3$ in the high density domain
prohibits the formation of the state with the same direction of
neutron and proton spins and leads to the appearance of the state
with the oppositely directed spins of neutrons and protons at high
densities. However, the results with the Gogny effective
interaction are in contrast with the results of microscopic
calculations with a realistic NN interaction~\cite{VB}, predicting
that the FM spin state is always preferable over the AFM one for
all relevant densities, but is less favorable compared to the
normal state.

\begin{figure}[tb]
\includegraphics[height=7.0cm,width=8.6cm,trim=48mm 142mm 57mm 69mm,
draft=false,clip]{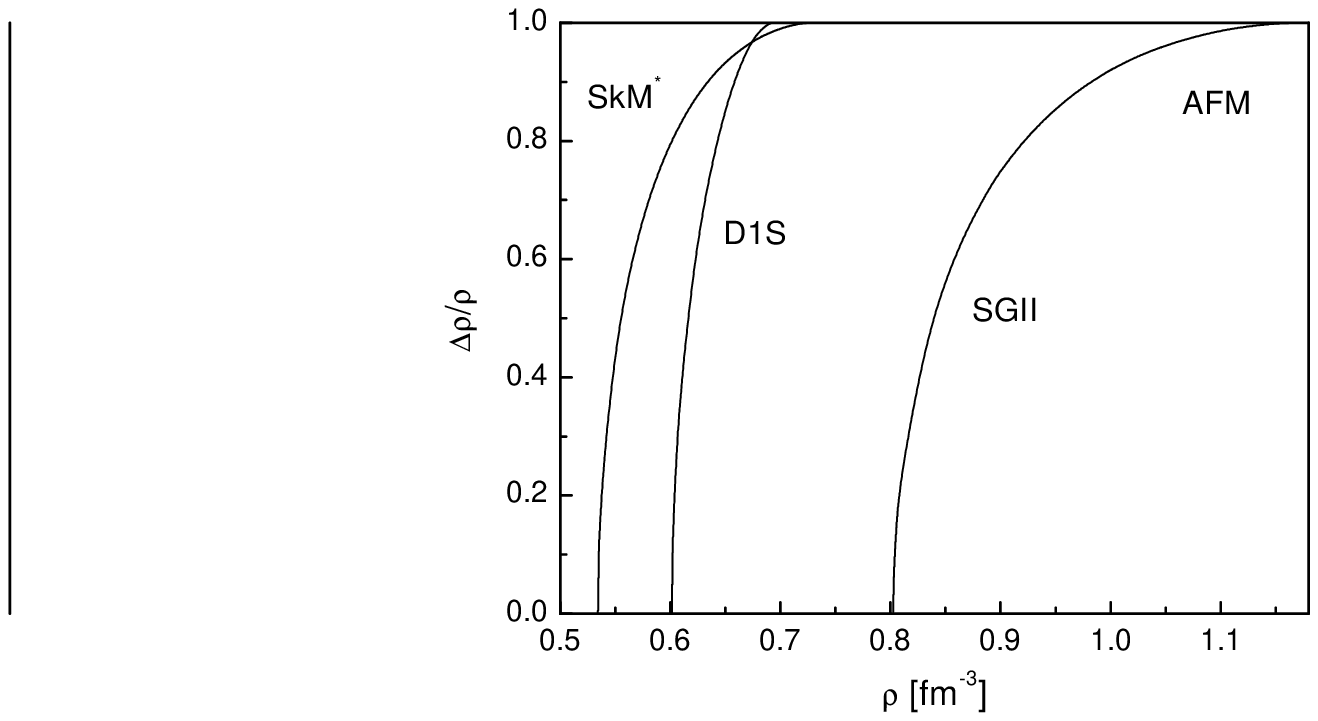}  \caption{AFM spin polarization
parameter as a function of density at zero temperature for the D1S
 Gogny force and  the SkM$^*$, SGII Skyrme forces. }\label{fig1}
\end{figure}

In Fig.~\ref{fig1} it is shown the dependence of the  AFM spin
polarization parameter
$\Delta\varrho_{\uparrow\downarrow}/\varrho$  as a function of
density at zero temperature. The AFM spin order parameter arises
at density $\varrho\approx3.8\varrho_0$ for the D1S potential. A
totally antiferromagnetically polarized state
($\Delta\varrho_{\uparrow\downarrow}/\varrho=1$) is formed at
$\varrho\approx4.3\varrho_0$. The neutron and proton spin
polarization parameters for the AFM spin ordered state are
opposite in sign and equal to
$$\Pi_n=-\Pi_p=\frac{\Delta\varrho_{\uparrow\downarrow}}{\varrho}.$$

For comparison, we plot in Fig.~\ref{fig1} the density dependence
of the AFM spin polarization parameter  for the Skyrme effective
forces SkM$^*$ and SGII \cite{I}. It is seen that the results with
the D1S potential are close to those with the SkM$^*$ potential
(for the D1S force the AFM spin polarization parameter is
saturated within a  narrower density domain than for the SkM$^*$
force).

\begin{figure}[bt]
\includegraphics[height=7.0cm,width=8.6cm,trim=48mm 142mm 54mm 69mm,
draft=false,clip]{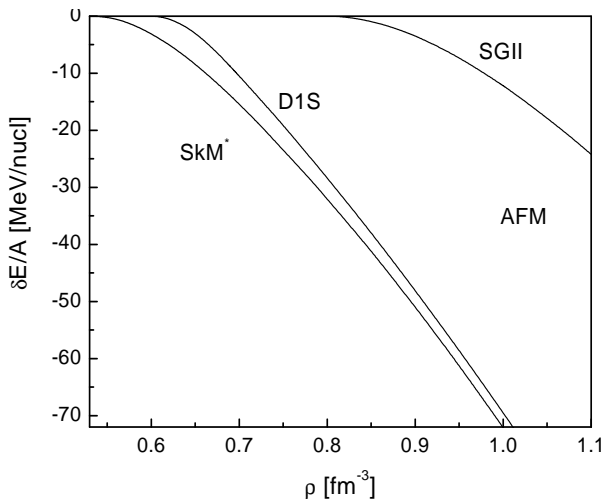}  \caption{ Total energy   per nucleon,
measured from its value in the normal state,  for the AFM spin
state  as a function of density at zero temperature for the D1S
 Gogny force and  the SkM$^*$, SGII Skyrme forces.}
 \label{fig2}
\end{figure}

To check the thermodynamic stability of the spin ordered state
with  oppositely directed spins of neutrons and protons, it is
necessary to compare the free energies of this state and the
normal state. In Fig.~\ref{fig2},  the difference between the
total energies per nucleon  of the spin ordered and normal states
is shown as a function of density at zero temperature. One can see
that nuclear matter in the model with  the D1S effective force
undergoes at some critical density a phase transition to the AFM
spin state.

\begin{figure}[bt]
\includegraphics[height=7.0cm,width=8.6cm,trim=48mm 142mm 54mm 69mm,
draft=false,clip]{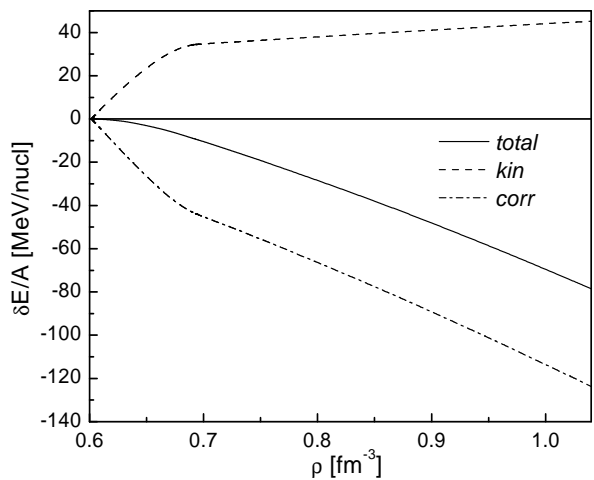}  \caption{ Total energy   per nucleon,
measured from its value in the normal state,  for the AFM spin
state  as a function of density at zero temperature for the D1S
 Gogny force, decomposed into  kinetic and correlation parts.}
 \label{fig3}
\end{figure}

In Fig.~\ref{fig3}, the difference between the total energies per
nucleon of the spin polarized and normal states is decomposed into
two parts, the kinetic and correlation ones. In spite of
increasing the kinetic energy per nucleon in the AFM spin state,
the AFM spin state becomes thermodynamically preferable over the
normal state due to the energy gain caused by medium correlations,
mainly by spin--isospin correlations, leading to AFM instability
of the ground state.

\section{Discussion and Conclusions}

Spin instability is a common feature associated with a large class
of Skyrme models, but is not realized in more microscopic
calculations. In this respect, it is interesting to study the
possibility of appearance of spin polarized states in nuclear
matter, utilizing another type of NN potential. Here we used the
finite range Gogny effective interaction, being successful in
describing nuclei and their excited states. The force parameters
are determined empirically by calculating the ground state in the
Hartree--Fock approximation and by fitting the observed ground
state properties of nuclei and nuclear matter. The analysis based
on the Gogny interaction shows that the self--consistent equations
of symmetric nuclear matter have solutions corresponding to  AFM
spin ordering and have no  solutions at all corresponding to  FM
spin ordering. This result is in contrast to  calculations with
the Skyrme interaction, predicting a FM phase transition in
symmetric nuclear matter~\cite{I}. In the last case,   the
SkM$^*$, SGII parametrizations were used, developed  to fit the
properties of nucleon systems with  small isospin asymmetry.
However, if the spin ordered state with  oppositely directed spins
of neutrons and protons for the Gogny interaction survives in the
domain of high isospin asymmetry, then this result can coincide
with the calculations for the Skyrme interaction, predicting the
appearance of this state in strongly asymmetric nuclear
matter~\cite{IY}. In the last case, the calculations were done
with the SLy4, SLy5 parametrizations, developed to reproduce the
properties of nuclear matter with high isospin asymmetry.

In a microscopic approach, one starts with the bare interaction
and obtains an effective particle--hole interaction by solving
iteratively the Bethe--Goldstone equation. In contrast to the
Skyrme and Gogny models, calculations with realistic NN potentials
predict more repulsive total energy per particle for a polarized
state comparing to the  nonpolarized one for all relevant
densities, and, hence, give no  indication of a phase transition
to a spin ordered state.

It must be emphasized that the interaction in the spin and isospin
dependent channels is a crucial ingredient in calculating spin
properties of nuclear matter and different behavior at high
densities of the interaction amplitudes, describing spin--spin and
spin--isospin correlations, lies behind this divergence in
calculations with  effective and realistic potentials, from one
side, and calculations with different types of effective forces,
from the other side. Since our calculations with the Gogny
interaction predict the AFM spin state as the  ground state of
symmetric nuclear matter, this emphasizes the role of
spin--isospin correlations in the high density region. Due to the
antiferromagnetic spin polarization, some neutrons and protons
with  opposite spins, e.g., spin up neutrons and spin down
protons, fill the Fermi surface with the larger radius and others,
spin down neutrons and spin up protons, occupy the Fermi surface
with the smaller radius. When  density increases, some neutrons
and protons  undergo spin--flip transitions from the inner Fermi
surface to the outer one due to increase  of spin--isospin
correlations. The usual way to constrain the interaction
parameters of spin--dependent amplitudes is based on the data on
isoscalar~\cite{T} and isovector (giant Gamow--Teller)~\cite{BDE}
spin--flip resonances. However, it is necessary to note that in
order to get  robust results for the spin polarization phenomena,
these constraints should be obtained for the high density region
of nuclear matter. Probably, such constraints can be obtained from
the data on the time decay of the magnetic field of isolated
neutron stars~\cite{PP}.

 In summary, we have considered the
possibility of phase transitions into spin ordered states of
symmetric nuclear matter within the Fermi liquid formalism, where
the NN interaction is described by the D1S Gogny effective force.
In contrast to the previous considerations, where the possibility
of formation of FM spin polarized states was studied on the basis
of calculation of the magnetic susceptibility,  we obtain
self--consistent equations for the FM and AFM spin order
parameters and solve them at zero temperature. It has been shown
in the model  with the D1S effective force that symmetric nuclear
matter undergoes a phase transition to the spin polarized state
with  oppositely directed spins of neutrons and protons, while the
state with the same direction of the neutron and proton spins does
not appear. The AFM spin order parameter arises at density
$\varrho\approx3.8\varrho_0$ and is saturated at
$\varrho\approx4.3\varrho_0$. These results may be of importance
for the adequate description of spin related phenomena in the
interior of neutron stars.

A.I. is grateful for  the support of the Topical Program of APCTP.

\end{document}